\documentclass[fleqn]{llncs}

\usepackage{amsmath,amssymb}
\usepackage{hyperref}

\hypersetup{%
  pdftitle={Computing the Kullback-Leibler Divergence between two Weibull Distributions},
  pdfauthor={Christian Bauckhage},
  pdfsubject={data science, statistics},
  pdfkeywords={Weibull distribution, Kullback-Leibler divergence},
  colorlinks=true,
  bookmarks=false}

\title{Computing the Kullback-Leibler Divergence between two Weibull Distributions}

\author{Christian Bauckhage}

\institute{%
  B-IT, University of Bonn, Bonn, Germany \\
  Fraunhofer IAIS, Sankt Augustin, Germany \\
  \email{http://mmprec.iais.fraunhofer.de/bauckhage.html}}

\begin{document}

\maketitle

\begin{abstract}
  We derive a closed form solution for the Kullback-Leibler divergence
  between two Weibull distributions. These notes are meant as
  reference material and intended to provide a guided tour towards a
  result that is often mentioned but seldom made explicit in the
  literature.
\end{abstract}

\section{The Weibull Distribution}

The Weibull distribution is the type III extreme value distribution;
its probability density function is defined for $x \in [0,\infty)$ and
given by
\begin{equation}
  \label{eq:Weibull}
  f(x \mid k, l) = \frac{k}{l} \left( \frac{x}{l} \right)^{k-1} \exp \left[- \left( \frac{x}{l} \right)^k \right]
\end{equation}
where $k > 0$ and $l > 0$ are shape and scale parameters,
respectively. This is a rather flexible, unimodal
density. Depending on the choice of $k$ and $l$, it may be skewed to
the left or to the right. For $k=1$, the Weibull coincides with the Exponential
distribution and for $k \approx 3.5$, it approaches the Normal
distribution.

A excellent account of the origins of the Weibull distribution is given in \cite{Rinne2008-TWD}.
Among others, it was introduced as a plausible
failure rate model \cite{Weibull1951-ASD} and has been frequently used for
life-time analysis in material- or actuary studies ever since. Extending its classical applications, it was
reported to account well for statistics of dwell times on Web sites
\cite{Liu2010-UWB}, times people spend playing online games
\cite{Bauckhage2012-HPL}, or the dynamics of collective attention on
the Web \cite{Bauckhage2013-MMO}. The Weibull also attracts interest
in machine learning or pattern recognition where it was found to
represent distributions of distances among feature vectors
\cite{Burghouts2007-TDF}, has been used in texture analysis
\cite{Geusebroek2005-ASS,Kwitt2008-ISM}, or was shown to provide a
continuous characterization of shortest paths distributions in random
networks \cite{Bauckhage2013-TWA}. Accordingly, methods for measuring
(dis)similarities of Weibull distributions are of practical interest
in data science for they facilitate model selection and statistical
inference.

\section{The Kullback-Leibler Divergence}

The Kullback-Leibler (KL) divergence provides a non-symmetric measure
of the similarity of two probability distributions $P$ and $Q$
\cite{Kullback1951-OIA}. In case both distributions are continuous, it
is defined as
\begin{equation}
  \label{eq:DKL}
  D_{KL} (P \parallel Q) = \int\limits_{-\infty}^{\infty} p(x) \, \log \frac{p(x)}{q(x)} \, dx
\end{equation}
where $p(x)$ and $q(x)$ denote the corresponding probability densities.

The KL divergence is a measure of relative entropy. It can be
understood as the information loss if $P$ is modeled by means of $Q$. Accordingly, the smaller
$D_{KL} (P \parallel Q)$, the more similar are $P$ and $Q$. Although
this is akin to the properties of a distance, the KL divergence does
not define a distance since it is neither symmetric nor satisfies the
triangle inequality.

\section{The KL Divergence between two Weibull Distributions}

Plugging two Weibull distributions $F_1$ and $F_2$ into \eqref{eq:DKL}
and noting once again that their densities are defined for $x \in [0,
\infty)$ immediately yields
\begin{equation}
  \label{eq:DKLWB}
  D_{KL} (F_1 \parallel F_2) = \int\limits_{0}^{\infty} f_1(x \mid k_1, l_1) \log \frac{f_1(x \mid k_1, l_1)}{f_2(x \mid k_2, l_2)} dx.
\end{equation}

\subsection{Step by Step Solution}

A somewhat obvious starting point for solving this expression is to
evaluate the logarithmic factor inside the integral. Given the
Weibull density in \eqref{eq:Weibull}, this factor
can be written as
\begin{equation*}
  \label{eq:log}
  \log \frac{k_1 / l_1 \left( x / l_1 \right)^{k_1-1} e^{- (x/l_1)^{k_1}}} {k_2 / l_2 \left( x / l_2 \right)^{k_2-1} e^{- (x/l_2)^{k_2}}}
  = \underbrace{\log \frac{k_1 / l_1}{k_2 / l_2  \vphantom{\left( x / l_2 \right)^{k_2-1}}}}_{A}
  + \underbrace{\log \frac{\left( x / l_1 \right)^{k_1-1}}{\left( x / l_2 \right)^{k_2-1}}}_{B}
  + \underbrace{\log \frac{e^{- (x/l_1)^{k_1}}}{e^{- (x/l_2)^{k_2}}  \vphantom{\left( x / l_2 \right)^{k_2-1}}}}_{C}.
\end{equation*}
Term $A$ in the expansion on the right hand side is a constant that
does not depend on $x$. Regarding term $B$, some straightforward
algebra reveals that it amounts to
\begin{align*}
  B
  & = (k_1 - 1) \log \frac{x}{l_1} - (k_2 - 1) \log \frac{x}{l_2} \\
  & = (k_1 - k_2) \log x + \underbrace{(k_2 - 1) \log l_2 - (k_1 - 1) \log l_1}_{D}
\end{align*}
where $D$ is yet another constant independent of $x$. For term $C$, it
is easy to see that
\begin{equation*}
  C = \left( \frac{x}{l_2} \right)^{k_2} - \left( \frac{x}{l_1} \right)^{k_1}.
\end{equation*}

Plugging the two constants $A$ and $D$ as well as the three terms that depend
on $x$ back into \eqref{eq:DKLWB} results in
\begin{align}
  & \int\limits_{0}^{\infty} f_1(x \mid k_1, l_1) \left[ A + (k_1 - k_2) \log x + D + \left( \frac{x}{l_2} \right)^{k_2} - \left( \frac{x}{l_1} \right)^{k_1} \right] dx \notag \\
  = & \; (A + D) \int\limits_{0}^{\infty} f_1(x \mid k_1, l_1) \, dx \label{eq:IntConst}\\
  & + \int\limits_{0}^{\infty} f_1(x \mid k_1, l_1) \, (k_1 - k_2) \log x \, dx \label{eq:IntLog}\\
  & + \int\limits_{0}^{\infty} f_1(x \mid k_1, l_1) \left( \frac{x}{l_2} \right)^{k_2} \, dx \label{eq:Intk2}\\
  & - \int\limits_{0}^{\infty} f_1(x \mid k_1, l_1) \left( \frac{x}{l_1} \right)^{k_1} \, dx \label{eq:Intk1}
\end{align}

Consider the integrals in \eqref{eq:IntConst} to \eqref{eq:Intk1} one
by one. First of all, since $f_1(x \mid k_1, l_1)$ is a probability
density function, the integral $\int_{0}^{\infty} f_1(x \mid k_1, l_1)
\, dx = 1$. The term in \eqref{eq:IntConst} therefore simplifies to
$A+D$.

Second of all, the integral in \eqref{eq:IntLog} can be solved using a
change of variables. In particular, consider the substitution
\begin{equation}
  \label{eq:subst}
  y = \left( \frac{x}{l_1}\right)^{k_1}
  \qquad
  \text{so that}
  \qquad
  dy = \frac{k_1}{l_1} \left( \frac{x}{l_1}\right)^{k_1-1} dx.
\end{equation}
Also, since $x = y^{1/k_1} l_1$, it follows that $\log x =
\tfrac{1}{k_1} \log y + \log l_1$. Together with the definition in
\eqref{eq:Weibull}, the term in \eqref{eq:IntLog} therefore becomes
\begin{align*}
  & (k_1 - k_2) \int\limits_{0}^{\infty} \frac{k_1}{l_1} \left( \frac{x}{l_1}\right)^{k_1 - 1} \, e^{-\left( x/l_1 \right)^{k_1}} \log x \; dx \\
  = & \; (k_1 - k_2) \int\limits_{0}^{\infty} e^{-y} \left[ \frac{1}{k_1} \log y + \log l_1 \right] \, dy \\
  = & \; (k_1 - k_2) \left[ \frac{1}{k_1} \int\limits_{0}^{\infty} e^{-y} \log y \; dy + \log l_1 \int\limits_{0}^{\infty} e^{-y} \, dy \right] \\
  = & \; (k_1 - k_2) \left[ - \frac{\gamma}{k_1} + \log l_1 \right].
\end{align*}
where $\gamma \approx 0.5772$ is the Euler-Mascheroni constant.

Third of all, in order to solve the integral in \eqref{eq:Intk2}, once again
consider the substitution in \eqref{eq:subst} and note that
$x^{k_2} = y^{k_2/k_1} l_1^{k_2}$. The term in \eqref{eq:Intk2} can
then be written as follows
\begin{align*}
  & \frac{1}{l_2^{k_2}} \int\limits_{0}^{\infty} \frac{k_1}{l_1} \left( \frac{x}{l_1}\right)^{k_1 - 1} \, e^{-\left( x/l_1 \right)^{k_1}} x^{k_2} \; dx \\
  = & \; \left( \frac{l_1}{l_2}\right)^{k_2} \int\limits_{0}^{\infty} e^{-y}  y^{k_2/k_1} \, dy \\
  = & \; \left( \frac{l_1}{l_2}\right)^{k_2} \Gamma \left(\frac{k_2}{k_1} + 1 \right).
\end{align*}
where $\Gamma (\cdot)$ is the gamma function.

Finally, fourth of all, noting that $x^{k_1} = y \, l_1^{k_1}$ immediately
allows for solving the term in \eqref{eq:Intk1}. It simply amounts to
\begin{align*}
  & - \frac{1}{l_1^{k_1}} \int\limits_{0}^{\infty} \frac{k_1}{l_1} \left( \frac{x}{l_1}\right)^{k_1 - 1} \, e^{-\left( x/l_1 \right)^{k_1}} x^{k_2} \; dx \\
  = & \; - \left( \frac{l_1}{l_1}\right)^{k_1} \int\limits_{0}^{\infty} e^{-y} \,  y \, dy \\
  = & \; -1.
\end{align*}

\subsection{Final Result}

Now, putting all intermediate results back together
establishes that: The KL divergence between two Weibull
  densities $f_1(x \mid k_1, l_1)$ and $f_2(x \mid k_2, l_2)$ amounts
  to
\begin{align*}
  & \log \frac{k_1 / l_1}{k_2 / l_2} + \log \frac{l_2^{k_2-1}}{l_1^{k_1-1}} + (k_1 - k_2) \left[ \log l_1 - \frac{\gamma}{k_1} \right] + \left( \frac{l_1}{l_2}\right)^{k_2} \Gamma \left(\frac{k_2}{k_1} + 1 \right) - 1 \\
  = & \; \log \frac{k_1}{l_1^{k_1}} - \log \frac{k_2}{l_2^{k_2}} + (k_1 - k_2) \left[ \log l_1 - \frac{\gamma}{k_1} \right] + \left( \frac{l_1}{l_2}\right)^{k_2} \Gamma \left(\frac{k_2}{k_1} + 1 \right) - 1.
\end{align*}

\section{Concluding Remarks}

Given the above results, it is instructive to verify it for a special
case. Setting the shape parameters $k_1$ and $k_2$ of two independent
Weibull distributions both to $1$ produces two Exponential
distributions with inverse rate parameters $1/l_1$ and $1/l_2$,
respectively. The closed form expression for the KL divergence
between the two distributions then simplifies to
\begin{equation*}
  \log l_2 - \log l_1 + \left( \frac{l_1}{l_2}\right) - 1
\end{equation*}
which indeed corresponds to the rather well known KL divergence between
two Exponential distributions.

Finally, given the above result, it is straightforward to compute symmetric divergence measures such as $\tfrac{1}{2} \left( D_{KL} (F_1 \parallel F_2) + D_{KL} (F_2 \parallel F_1) \right)$ and use these to define kernel functions, for instance, using the method discussed in \cite{Moreno2003-AKL}.

\bibliographystyle{splncs}
\bibliography{literature}

\end{document}